\documentclass[%
reprint,
superscriptaddress,
amsmath,amssymb,
aps,
pra,
]{revtex4-2}

\usepackage[dvipsnames]{xcolor}
\usepackage[utf8]{inputenc}
\usepackage[english]{babel}
\usepackage{amsmath}
\usepackage{graphicx}  
\usepackage{dcolumn} 
\usepackage{bm} 
\usepackage[hidelinks]{hyperref} 
\usepackage[detect-weight=true,separate-uncertainty=true]{siunitx}  
\usepackage{svg}
\usepackage{soul}
\usepackage{newfloat}
\usepackage[all]{hypcap}
\usepackage[capitalise,nameinlink]{cleveref}

\graphicspath{{figs/}}  

\definecolor{bluegray}{RGB}{40,180,160}
\definecolor{navygray}{RGB}{110,140,170}
\hypersetup{
citecolor={bluegray},
linkcolor={navygray},
urlcolor={bluegray},
}

\DeclareFloatingEnvironment[name={Figure S.\!\!}]{suppfigure}
\DeclareFloatingEnvironment[name={Table S.\!\!}]{supptable}

\crefname{suppfigure}{Fig. S.\!\!}{Figs. S.\!\!}
\Crefname{suppfigure}{Figure S.\!\!}{Figures S.\!\!}

\crefname{supptable}{Tab. S.\!\!}{Tabs. S.\!\!}
\Crefname{supptable}{Table S.\!\!}{Tables S.\!\!}

\newcommand{\bd}[1]{{\color{Black}{#1}}}

\usepackage{epstopdf}
\usepackage{braket}
\usepackage{upgreek}
\usepackage{amsmath}
\usepackage{siunitx}
\usepackage{changes}
\usepackage{natbib}
\usepackage[title]{appendix}

\usepackage{multirow}

\begin{document}

\title{Offset Charge Dependence of Measurement-Induced Transitions in Transmons}

\author{Mathieu~F\'echant}
\email{mathieu.fechant@kit.edu}
\affiliation{IQMT,~Karlsruhe~Institute~of~Technology,~76131~Karslruhe,~Germany}

\author{Marie Fr\'ed\'erique Dumas}
\email{Marie.Frederique.Dumas@USherbrooke.ca}
\affiliation{Institut Quantique and D\'epartement de Physique, Universit\'e de Sherbrooke, Sherbrooke J1K 2R1 QC, Canada}

\author{Denis~B\'en\^atre}
\affiliation{IQMT,~Karlsruhe~Institute~of~Technology,~76131~Karslruhe,~Germany}

\author{Nicolas~Gosling}
\affiliation{IQMT,~Karlsruhe~Institute~of~Technology,~76131~Karslruhe,~Germany}

\author{Philipp~Lenhard}
\affiliation{IQMT,~Karlsruhe~Institute~of~Technology,~76131~Karslruhe,~Germany}

\author{Martin~Spiecker}
\affiliation{IQMT,~Karlsruhe~Institute~of~Technology,~76131~Karslruhe,~Germany}

\author{\bd{Simon~Geisert}}
\affiliation{PHI,~Karlsruhe~Institute~of~Technology,~76131~Karlsruhe,~Germany}

\author{\bd{S{\"o}ren~Ihssen}}
\affiliation{PHI,~Karlsruhe~Institute~of~Technology,~76131~Karlsruhe,~Germany}

\author{Wolfgang~Wernsdorfer}
\affiliation{IQMT,~Karlsruhe~Institute~of~Technology,~76131~Karslruhe,~Germany}
\affiliation{PHI,~Karlsruhe~Institute~of~Technology,~76131~Karlsruhe,~Germany}

\author{Benjamin~D'Anjou}
\affiliation{Institut Quantique and D\'epartement de Physique, Universit\'e de Sherbrooke, Sherbrooke J1K 2R1 QC, Canada}

\author{Alexandre~Blais}
\affiliation{Institut Quantique and D\'epartement de Physique, Universit\'e de Sherbrooke, Sherbrooke J1K 2R1 QC, Canada}
\affiliation{CIFAR, Toronto, M5G 1M1 Ontario, Canada}

\author{Ioan~M.~Pop}
\email{ioan.pop@kit.edu}
\affiliation{IQMT,~Karlsruhe~Institute~of~Technology,~76131~Karslruhe,~Germany}
\affiliation{PHI,~Karlsruhe~Institute~of~Technology,~76131~Karlsruhe,~Germany}
\affiliation{Physics~Institute~1,~Stuttgart~University,~70569~Stuttgart,~Germany}

\date{\today}

\begin{abstract}
A key challenge in achieving scalable fault tolerance in superconducting quantum processors is readout fidelity, which lags behind one- and two-qubit gate fidelity. A major limitation in improving qubit readout is measurement-induced transitions, also referred to as qubit ionization, caused by multiphoton qubit-resonator excitation occurring at specific photon numbers. Since ionization can involve highly excited states, it has been predicted that in transmons---the most widely used superconducting qubit---the photon number at which measurement-induced transitions occur is gate charge dependent. This dependence is expected to persist deep in the transmon regime where the qubit frequency is gate charge insensitive. We experimentally confirm this prediction by characterizing measurement-induced transitions with increasing resonator photon population while actively \bd{calibrating} the transmon's gate charge. Furthermore, because highly excited states are involved, achieving quantitative agreement between theory and experiment requires accounting for higher-order harmonics in the transmon Hamiltonian.
\end{abstract}


\maketitle 


Circuit quantum electrodynamics (cQED) with transmon qubits is a leading platform for quantum information processing, enabling dispersive qubit readout via coupling to a microwave resonator~\cite{Blais2004cQED,Wallraff2005,Koch2007ChargeInsensitive}. 
Impressive progress has been achieved towards high-fidelity and quantum non-demolition (QND) qubit readout in this architecture, notably thanks to the development of amplifiers operating near the quantum limit~\cite{Yamamoto2008, Castellanos-Beltran2008, PhysRevLett.106.110502,
Macklin2015Sep} and to \bd{device optimization}~\cite{Bultink_Dicarlo_2016,PhysRevApplied.7.054020, Takmakov2021Jun, Sunada2022Apr, Swiadek2024Enhancing, jerger2024dispersivequbitreadoutintrinsic, Spring2024}. A key tenet of the dispersive readout is that increasing the number of photons probing the readout resonator should improve signal-to-noise ratio (SNR) while preserving QND~\cite{Blais2004cQED}. However, it is experimentally observed that increasing the photon number leads to unwanted qubit transitions, thereby negating the benefits of strong readout drives~\cite{Jeffrey2014,Sank2016Measurement, Minev2019, PhysRevApplied.7.054020, Khezri2023MIST}. This limits the rate of information extraction, creating a bottleneck for error correction in superconducting quantum processors.

Measurement-induced transitions \bd{into high-energy levels of the transmon} have been attributed to multiphoton resonances occurring at specific intraresonator photon numbers~\cite{Sank2016Measurement}. This observation \bd{has led to a theoretical framework} for understanding this phenomenon---referred to as measurement-induced transitions (MIST) and ionization in the literature---with predictions that are in good agreement with experiments~\cite{Sank2016Measurement,Shillito2022Dynamics,Khezri2023MIST,Cohen2023chaos,Xiao2023Diagrammatic,dumas2024unified}. Crucially, because they involve high-energy states of the transmon, these resonances, and their associated critical photon numbers, have been predicted to be gate-charge dependent~\cite{Cohen2023chaos,dumas2024unified}. This stands in contrast to the transmon's 0-1 transition frequency, whose gate-charge dependence is exponentially suppressed with increasing ratio of the qubit's Josephson energy $E_J$ to charging energy $E_C$~\cite{Koch2007ChargeInsensitive}. Moreover, because they affect high-energy states, higher-order harmonics of the transmon Hamiltonian~\cite{Willsch2024higherharmonics} are expected to influence its ionization. 

In this work, we present experimental observations confirming the role of gate charge and higher-order harmonics on measurement-induced state transitions. To this end, we measure the impact of the resonator photon population on the qubit state as a function of the average photon number $\bar{n}_r$ and of the qubit frequency $\omega_{01}$ for two transmons of different $E_J/E_C$ ratios. A previous experiment indirectly probed the gate-charge dependence of ionization by observing shot-to-shot variations in the critical photon number that were attributed to gate charge fluctuations~\cite{Khezri2023MIST}. Here, the gate charge $n_g$ is actively \bd{calibrated}, allowing us to directly confirm the \bd{theory}~\cite{dumas2024unified}. This understanding allows us to identify robust regions for readout as a function of $n_g$, and will inform future qubit calibrations, optimal control, and design strategies. 

\begin{figure}[!t]
\includegraphics[width = 0.9\columnwidth]{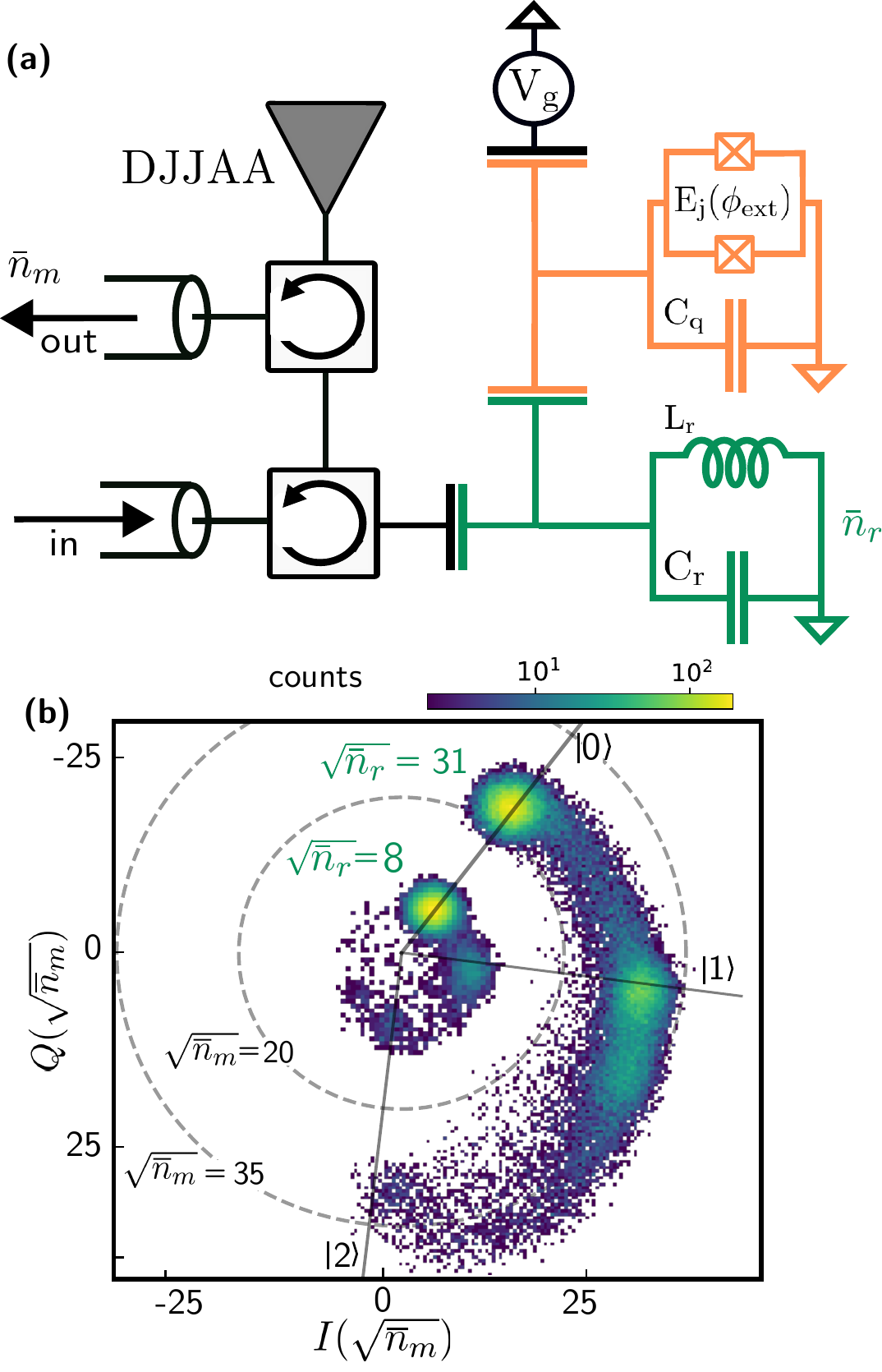}
\caption{ \textbf{Dispersive readout. } \textbf{(a)} Schematic qubit-resonator setup with \bd{active charge calibration.} The magnetic flux ($\phi_{\mathrm{ext}}$) tunable transmon (orange) is capacitively coupled to a readout resonator (green) measured in reflection through a \bd{Josephson} amplifier~\cite{DJJA_ref}. See Ref.~\cite{Supplemental}
for the full experimental setup. The qubit is capacitively coupled to a line that allows changing the charge offset $n_g$. \textbf{(b)} \bd{$IQ$} scatter plot of the dispersive measurement outcomes of transmon A. We continuously pump the readout resonator at \bd{frequency} $\omega_d/2\pi = \SI{6.11972}{GHz}$ and integrate the output every 2 $\mathrm{\mu s}$. We show the resulting histograms for two different experiments using two different resonator photon numbers, $\sqrt{\bar{n}_r }= 8$ and $\sqrt{\bar{n}_r }= 31$.
}
\label{fig_design}
\end{figure}

We use a standard cQED setup consisting of a flux tunable transmon coupled to a readout resonator measured in reflection; see Fig.~\ref{fig_design}(a). The transmon is capacitively coupled to a line which allows microwave drive and dc charge bias. We apply a readout drive at the resonator input port, loading $\bar{n}_r$ photons. The reflected signal undergoes amplification and we report the measured $I$ and $Q$ quadratures; see Fig.~\ref{fig_design}(b). Here and below, this is reported in units of the measurement photon number $\bar{n}_m = \bar{n}_r \kappa T_m /4$ during the integration time $T_m$~\cite{n_meas_0}, where $\kappa$ is the resonator damping rate. The measured values cluster around several $IQ$ coordinates, each corresponding to a transmon state. Deviations from non-QND behavior are evident from the appearance of clusters away from \bd{that of} the initial qubit state, here $\ket0$.

\begin{figure}[!t]
\includegraphics[width = \columnwidth]{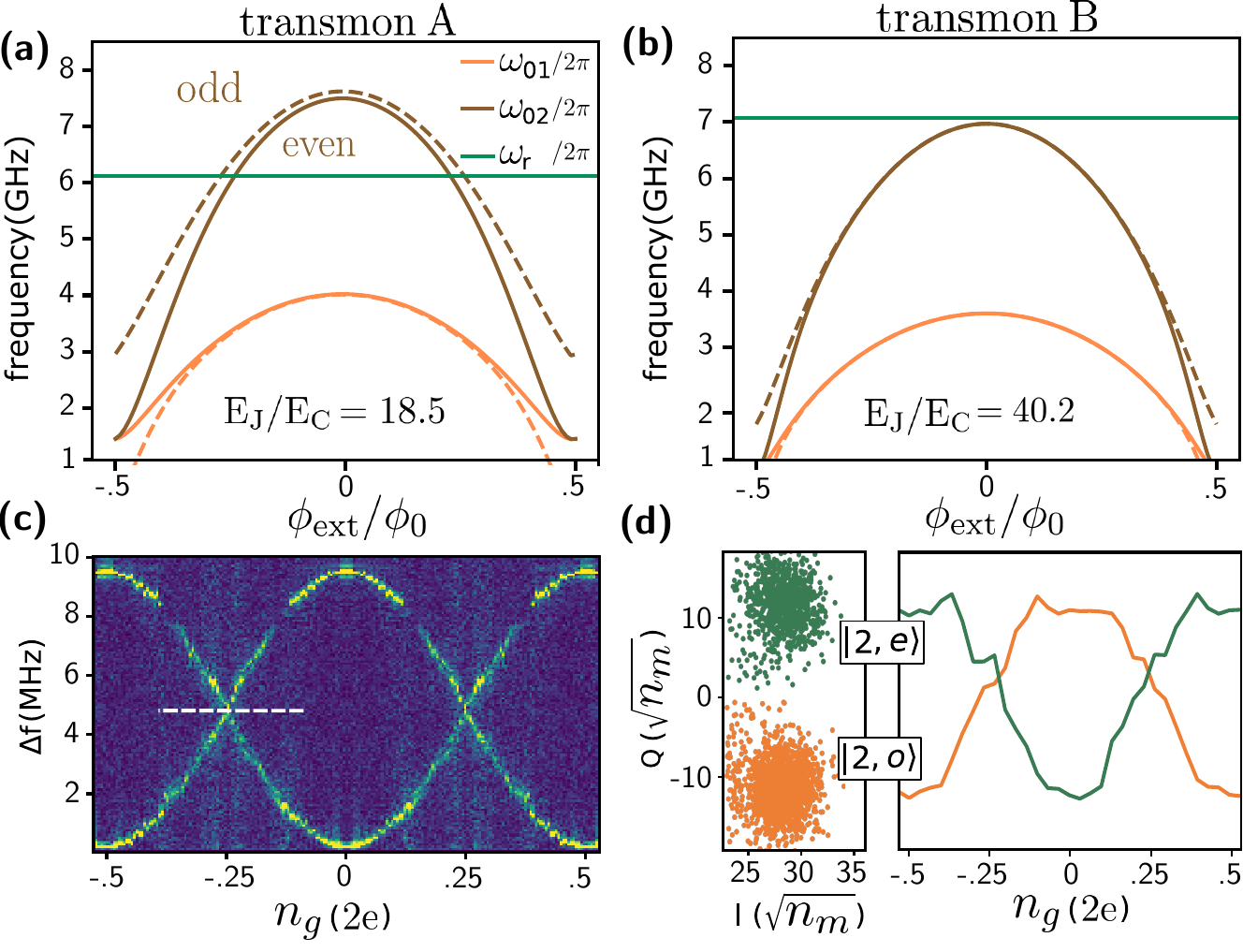}
\caption{\textbf{Flux and Charge Dependence.} \textbf{(a,b)} Flux dependence of the 0-1 and 0-2 transmon transitions for device A ($E_J/E_C = 18.5$) and device B ($E_J/E_C = 40.2$) at $n_g=0$. Energy levels are shown for both even parity states (full lines) and odd parity states (dashed lines) assuming symmetric junctions. \textbf{(c)} Charge dependence of the Fourier transform of a Ramsey interference experiment performed at \bd{frequency} 3.9965 GHz on device A, with $\mathrm{\Delta f}$ the frequency difference to the Ramsey pulse. The dashed line indicates the average $\bar{f}_{01}=3.9992\,\mathrm{GHz}$. \textbf{(d)} Left panel: $IQ$ clouds for states $|2,\mathrm{o}\rangle $ and $|2,\mathrm{e}\rangle$ after a 4 $\mathrm{\mu s}$ pulse for device B. Right panel: Imaginary part of the even and odd second transmon excited state distributions over one charge period.} 
\label{fig_transmons}
\end{figure}

To characterize the measurement-induced transitions as a function of external flux and gate charge offset, we first use a device (device A) with a readout resonator frequency $\omega_r/2\pi=\SI{6.12}{GHz}$ and decay rate \bd{$\kappa/2\pi = \SI{.38}{MHz}$}.
The resonator is coupled with strength $g/2\pi=\SI{13}{MHz}$ to a transmon qubit of charging energy $E_C/2\pi = \SI{365}{MHz}$ and maximum Josephson energy $E_J/2\pi = \SI{6.71}{GHz}$ at zero flux bias $\phi_{\mathrm{ext}}=0$. With a maximum $E_J/E_C$ ratio of $\sim 18.5$, this device is in the shallow transmon regime with $\sim \SI{9}{MHz}$ charge dispersion of the 0-1 transition \bd{and $T_1 \approx \SI{30}{\mu s}$ at the sweet spot}.
\Cref{fig_transmons}(a) shows the flux dependence of the transmon's transition frequencies between the ground state and the first two excited states. \bd{Our calibration of the charge offset relies on measuring the} Ramsey fringes of the 0-1 transition as a function of the \bd{applied offset voltage, which reveals} two sinusoids of periodicity $2e$; see \cref{fig_transmons}(c)~~\cite{serniak_hot_2018,Rist__2013}. The two measured frequencies result from random quasiparticle tunneling events shifting the response by $1e$; see also the full and dashed lines in \cref{fig_transmons}(a) labeled even and odd, respectively~\cite{Schreier2008}. \bd{Before an experiment, we measure the frequency at a few offset voltages, which takes 10 s. A sinusoidal fit yields the offset voltage for which $n_g=0$. We then set $n_g$ by adjusting the offset voltage relative to $n_g=0$. We repeat this procedure every minute, setting $n_g$ with $2\%$ precision.}

\begin{figure*}
\hspace*{0cm}
\includegraphics[width = 2\columnwidth]{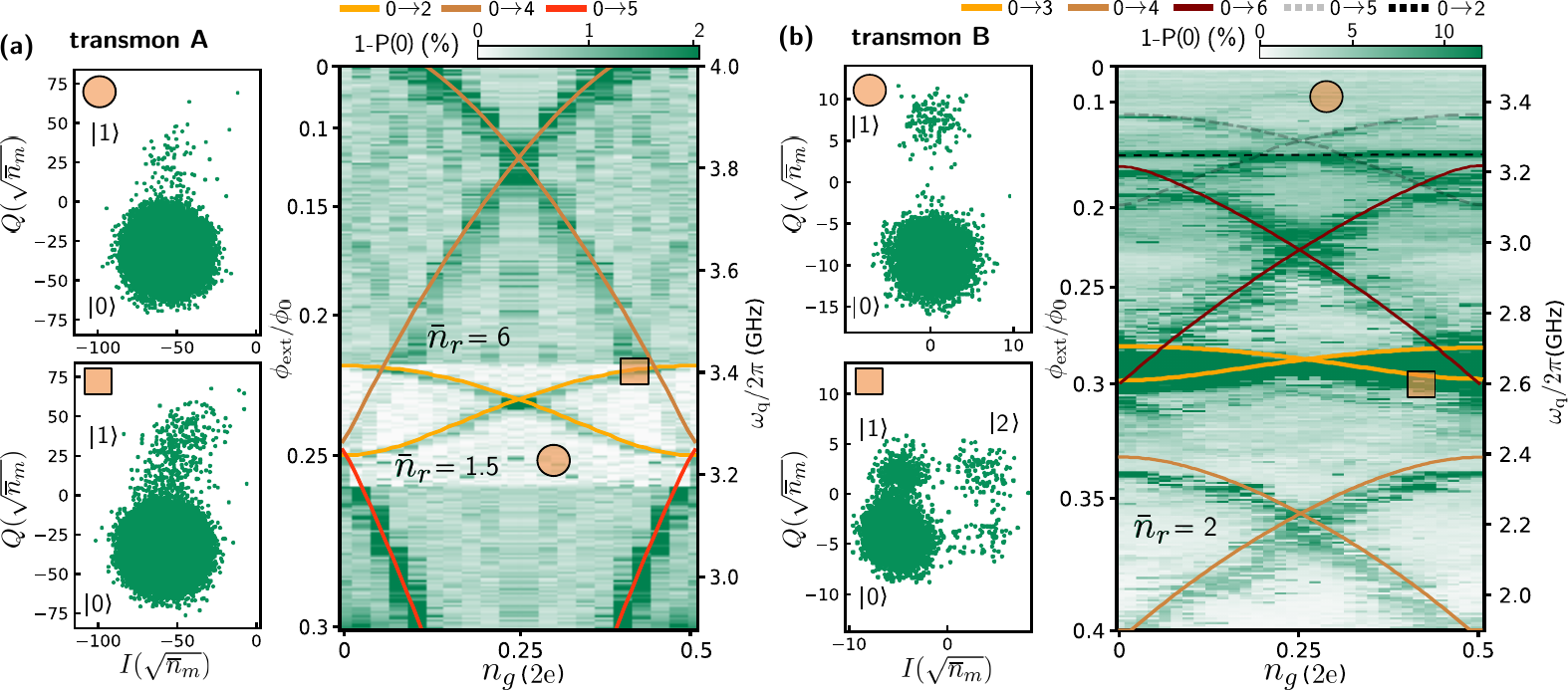}
\caption{ \textbf{ Probability $\boldsymbol{1-P(0)}$ to find the transmon in an excited state vs. flux and charge offset.} \textbf{(a)} For device A, we continuously populate the resonator with $\bar{n}_r \approx 6$ photons at frequency $\omega_d/2\pi=\SI{6.11972}{GHz}$ and integrate over 25 $\mu s$. 
In the central part of the plot, we lower the photon number to $\bar{n}_r \approx 1.5$ to reduce the width of the features.
\textbf{(b)} For device B, we stroboscopically pump the resonator with $\bar n_r \approx 2 $ photons at \bd{frequency} $\omega_d/2\pi=\SI{7.0535}{GHz}$ with a 2 $\mathrm{\mu s}$ pulse every 3 $\mathrm{\mu s}$. In both panels, the qubit frequency corresponding to $\phi_{\mathrm{ext}}$ and $n_{g} =0$ is indicated by the right axis. \bd{The side panels show $IQ$ clouds ($10^5$ shots) for selected values of flux and gate charge to highlight the contrast between negligible (circle) and significant (square) leakage. Residual leakage does not seem to be limited by qubit temperature but by other sources. The photon number $\bar{n}_{r}$ is calibrated with a low-power ac-Stark shift experiment.} The multiphoton resonance conditions $\omega_{0j}=n\omega_d$ (labeled $0\to j$) are plotted on top of the experimental results (orange-red lines). The remaining discrepancies $\lesssim \SI{100}{MHz}$ are consistent with \bd{corrections not included in} our model, such as junction asymmetry. The dashed lines indicate the theory for inelastic scattering with a spurious mode of frequency \bd{$\omega_s = 2\pi \times \SI{0.78}{GHz}\,({\rm mod}\,\omega_d)$. }
}
\label{fig_matchings}
\end{figure*}

To confirm the importance of gate charge on ionization deeper in the transmon regime, where the computational states have a much weaker dependence on gate charge, we also measure a device (device B) with a charging energy $E_C/2\pi=\SI{217}{MHz}$ and maximum Josephson energy $E_J/2\pi=\SI{8.72}{GHz}$ at zero flux bias $\phi_{\mathrm{ext}}=0$, yielding $E_J(\phi_{\mathrm{ext}})/E_C \leq 40.2$; see \cref{fig_transmons}(b). This qubit is coupled with strength $g/2\pi=\SI{186.5}{MHz}$ to a readout resonator of frequency $\omega_r/2\pi=\SI{7.05}{GHz}$ and decay rate $\kappa/2\pi=\SI{0.92}{MHz}$. \bd{We measure $T_1 \approx $\SI{50}{\micro s} at the sweet spot.} At this large $E_J/E_C$ ratio, the charge dispersion of $\sim \SI{50}{kHz}$ is too small to be resolved through Ramsey interferometry. To calibrate the gate charge, we instead \bd{monitor the charge offset imprinted on the resonator's dispersive shift for state $|2\rangle$~\cite{serniak_direct_2019}; see Fig.~\ref{fig_transmons}(d). \bd{This sets} $n_g$ with better than $5\%$ precision.}

To map the measurement-induced transitions as a function of the qubit control parameters, we monitor the qubit state by probing the resonator response with a maximum of $\bar{n}_r \sim 6$ photons. \bd{Here, the resonator is pumped and probed continuously to avoid waiting for the long $\SI{2.6  }{\mathrm{\mu s}}$ decay time of the resonator.} The resulting probability to find the transmon in a state other than $\ket0$, $1-P(0)$, is reported in \cref{fig_matchings}. For both devices we observe flux- and gate-charge-dependent features symmetric about $n_g=0.25$ due to frequent parity switching induced by quasiparticle tunneling events. These features correspond to regions where transitions out of the ground state are more pronounced. The side panels show the resonator response in the $IQ$ plane on top of (square) and away from (circle) one of these features. Here, the moderate value of $\bar{n}_r \lesssim 6$ is chosen to avoid excessive broadening of the gate-charge-dependent features in the main panels and, as discussed below, to limit the qubit's ac-Stark shift. 

To understand these features, we model the field in the resonator as an effective classical drive on the transmon. \bd{The Hamiltonian is}~\cite{Cohen2023chaos,Lledo2023Cloaking,dumas2024unified}
\begin{align}
\hat H(t) = \hat H_{t} + \varepsilon_t(t) \cos(\omega_d t) \; \hat{n}_t,
\label{eq:H_driven_transmon}
\end{align} 
where $\hat H_{t}$ is the undriven transmon Hamiltonian. Here, we account for higher-order harmonics of the potential such that $\hat H_{t}$ reads~\cite{Willsch2024higherharmonics} 
\begin{align}
    \hat H_{t}=4E_C(\hat{n}_t-n_g)^2-\sum_{m\geq 1} E_{Jm}\cos(m\hat \varphi_t).
    \label{eq:H_transmon_main}
\end{align}
In this expression, $\hat{n}_t$ and $\hat\varphi_t $ are the transmon charge and phase operators, respectively, $\omega_d\approx\omega_r$ is the drive frequency, and $\varepsilon_t(t)=2g\sqrt{\bar{n}_r(t)}$ is the effective time-dependent drive amplitude~\cite{Supplemental}.
The charging energy $E_C$ and the Josephson energies $E_{Jm}$ are fitted to independently measured transition frequencies at different values of $n_g$~\cite{Supplemental}.
\bd{The higher harmonics are only fitted for device B since they mostly affect the offset charge dependence of the critical photon number, which is not probed for device A.}

\begin{figure*}[!t]
\hspace*{0cm}
\includegraphics[width = 2\columnwidth]{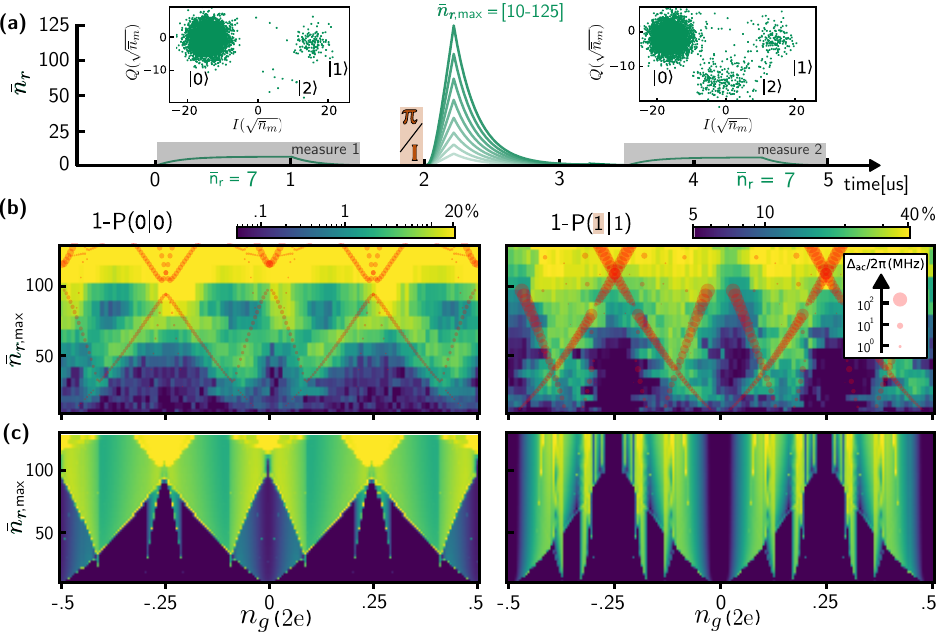} 
\caption{ \textbf{Probability of leaving the initial state for device B.} \textbf{(a)} Experimental pulse sequence \bd{performed at the flux sweet spot ($f_{01} = \SI{3.604}{GHz}$)}. We apply a high-power 200 ns readout pulse with variable amplitude ($\bar{n}_{r,\mathrm{max}} \in [10,125]$), straddled by two low-power 1 $\mathrm{\mu}s$ readout pulses ($\bar{n}_{r,\mathrm{max}} = 7$) for high-fidelity preparation and readout. An optional $\pi$-pulse enables excited state preparation. All pulses have frequency $\omega_d/2\pi=\SI{7.0535}{GHz}$. The insets show the $IQ$ data for the preparation and final measurements for $\bar{n}_{r,\mathrm{max}} = 50$ and $n_{g} = 0$. \textbf{(b)} Measurement-induced transition probability as a function of gate charge and maximum average resonator photon number when initializing the qubit in the ground state (left) or excited state (right). The photon number at higher powers is calibrated by extrapolating a nonlinear semiclassical model of resonator dynamics~\cite{Supplemental}.
Red circles indicate avoided crossings in the Floquet quasienergy spectrum. The dot area is proportional to the gap size $\Delta_{ac}$. (c) Numerical simulation of the experiment from the semiclassical time dynamics.
}
\label{fig:readout_demolition_comparison}
\end{figure*}

At low photon number $\bar{n}_r$, the qubit's ac-Stark shift is small and, following \cref{eq:H_driven_transmon}, we expect multiphoton transitions to occur when $\omega_{ij}\approx n \omega_{d}$, where $\omega_{ij} = \omega_j-\omega_i$ with $\omega_i$ a bare eigenfrequency of $\hat H_{t}$ and $n$ an integer corresponding to the number of readout photons involved in the process. The lines shown in \cref{fig_matchings} indicate the predicted resonance conditions assuming no junction assymetry for selected $i \rightarrow j$ transitions, as specified in the legend, and show remarkable agreement with the measured leakage probability. 
For device A, the features close to $\phi_\mathrm{ext}/\phi_0 = 0.23$ (light orange lines) correspond to a $0\rightarrow2$ transition involving a single drive photon, $\omega_{02}\approx\omega_d$; see \cref{fig_transmons}(a) where this resonance and its charge dispersion is also evident. Because this is a first-order process, non-QND behavior is very pronounced. For this reason, a smaller resonator photon number ($\bar{n}_r = 1.5$) is used in the vicinity of this resonance compared to the rest of the plot ($\bar{n}_r = 6$). For device B, a similar first-order resonance between the transmon states 0 and 3 with strong non-QND behavior is also observed (light orange lines).

\bd{Device B} also shows a large leakage probability around $\omega_{01}/2\pi=\SI{3.26}{GHz}$ for all values of $n_g$ that does not directly match a $0\to j$ multiphoton transition. This leakage can be explained by inelastic scattering of readout photons via a spurious mode \bd{at frequency $\omega_s=2\pi \times \SI{0.78}{GHz}\,({\rm mod} \, \omega_d)$} in the qubit environment~\cite{singh2025impactjosephsonjunctionarray,bista2025readoutinducedleakagefluxoniumqubit,Hoyau2025spurious,WeiDai}, for which the resonance condition \bd{$\omega_{02}+\omega_s \approx n \omega_d$} is satisfied \bd{for some integer $n$} (black dashed line). Assuming the existence of a mode at this frequency also predicts the increased leakage observed around \bd{$\omega_{05}+\omega_s\approx (n+1)\omega_d$} (gray dashed line). Away from the resonances, the residual transition probability shows a trend towards a more QND behavior as the qubit-resonator detuning increases, consistent with recent results~\cite{Kurilovich2025LargeDetunings}. \bd{This trend is not seen in device A because of a worse $IQ$ contrast and of a smaller probed frequency range than in device B.}

At larger resonator photon number, the transmon levels can be significantly ac-Stark shifted such that the multiphoton resonance conditions now involve the transmon frequencies dressed by the drive rather than the bare ones~\cite{Sank2016Measurement,Xiao2023Diagrammatic,dumas2024unified}. To measure ionization in this situation, we follow the measurement protocol shown in \cref{fig:readout_demolition_comparison}(a). We first prepare the transmon of device B, operated at the flux sweet spot, in state $|0\rangle$ by postselecting on the result of a first low-power \bd{QND} measurement ($\bar n_r\sim 7$). In half of the realizations, we then apply a $\pi$-pulse to prepare the excited state $|1\rangle$. Next, we populate the resonator with up to $\bar{n}_r=125$ photons. Finally, we assess the non-QND character of this strong drive by performing a second QND measurement to determine the qubit’s final state. As can be seen by comparing the two insets in \cref{fig:readout_demolition_comparison}(a), the strong drive results in population transfer to excited states. 

Figure~\ref{fig:readout_demolition_comparison}(b) shows the measured population transfer when starting in $|0\rangle$ (left) and $|1\rangle$ (right) as a function gate charge and resonator photon number. We observe a rich charge-dependent structure, with sharp increases in non-QNDness at specific $n_g$-dependent photon numbers. \bd{Importantly, we observe values of $n_g$ where QNDness persists up to much larger photon numbers, showing that our active charge calibration can mitigate measurement-induced transitions.}

To quantitatively understand these observations, we compute the \bd{exact} Floquet quasienergy spectrum of \cref{eq:H_driven_transmon} as a function of effective drive amplitude $\varepsilon_t$ on the qubit~\cite{Supplemental}.
From these quasienergies, which encapsulate the drive-induced ac-Stark shifts, we identify avoided crossings corresponding to multiphoton resonances, here shown as red dots in \cref{fig:readout_demolition_comparison}(b)~\cite{dumas2024unified}. The gap $\Delta_{\rm ac}$ at the avoided crossing, which is indicated by the dot area, increases with the effective drive amplitude, reflecting a stronger hybridization of the transmon with the drive. Importantly, the quantitative agreement between experimental results and the Floquet calculations seen in \cref{fig:readout_demolition_comparison}(b) is only obtained when including higher-order harmonics up to $m = 3$~\cite{Supplemental}.
This is because the observed transitions involve highly excited states that lie above the top of the cosine potential well. These states are strongly sensitive to higher-order harmonics and to the gate charge~\cite{Supplemental}.
However, this dependence of the critical photon number on higher-order harmonics does not provide sufficient information to determine the specific origin of these harmonics in our experiment~\cite{Supplemental}.

In \cref{fig:readout_demolition_comparison}(b), the QNDness does not decrease monotonically with increasing $\bar{n}_{r,\mathrm{max}}$; in some regions above a resonance, higher QNDness is observed. This behavior, consistent with the findings of \textcite{Sank2016Measurement}, arises from Landau-Zener transitions \bd{\cite{Grifoni1998}} that occur as the system sweeps through multiphoton resonances~\cite{Shillito2022Dynamics,dumas2024unified}. The resulting non-QNDness thus depends on both the rate at which a given resonance is traversed and the size of the associated energy gap $\Delta_\mathrm{ac}$ \bd{\cite{Breuer1989Adiabatic, Drese1999Floquet,Shevchenko2010LandauZenerStuckelberg,Ikeda2022Floquet-Landau-Zener}}. Because larger $\bar n_{r,\mathrm{max}}$ mean a faster crossing of resonances during the transients, a resonance that leads to non-QND behavior at small $\bar{n}_{r,\mathrm{max}}$ may no longer contribute at larger values. To model these complex dynamics, we solve the Schrödinger equation with the Hamiltonian of \cref{eq:H_driven_transmon} 
and following the same protocol as the experiment~\cite{Supplemental}.
The resulting theoretical transition probabilities are shown in \cref{fig:readout_demolition_comparison}(c). Crucially, the \bd{simulation accounts} for the rise and fall of the resonator population, which results in some resonances being traversed twice~\cite{Supplemental}.
Despite the \bd{model's simplicity}, we find remarkable agreement between \bd{experiment and theory}, without the use of adjustable parameters.

In summary, we have directly probed the gate charge dependence of measurement-induced transitions in transmons, confirming recent theoretical predictions~\cite{Cohen2023chaos,dumas2024unified}. This was made possible by active \bd{gate charge calibration}. A key finding is that achieving quantitative agreement between experiment and theory requires accounting for higher-order harmonics of the transmon Hamiltonian. Additionally, our results show that the ring-up and ring-down transients influence measurement-induced state transitions. 
Our findings \bd{demonstrate} that \bd{active charge calibration} can help avoid regions that are most susceptible to unwanted multiphoton transitions, therefore enabling a path towards higher fidelity QND readout. These results are broadly applicable to other nonlinear driven superconducting circuits dispersive readout, such as parametric gates and couplers, qubit reset protocols, and quantum state stabilization schemes. 

\section*{Acknowledgements}
We are grateful to Dennis Willsch, Madita Willsch, Alexandru Petrescu and Dennis Rieger for stimulating discussions. 
We thank L. Radtke and S. Diewald for technical assistance.
We thank Immanuel Speitelsbach for providing the Josephson amplifier
This work was financed by the German Ministry of Education and Research (BMBF) within project QSolid (FKZ:13N16151).
D.B.~acknowledges funding from the Horizon Europe program via Project No.~101113946 OpenSuperQPlus100.
N.G. and M.S.~acknowledge funding from the German Ministry of Education and Research (BMBF) within project GEQCOS (FKZ:~13N15683).
Facilities use was supported by the KIT Nanostructure Service Laboratory. 
We acknowledge the measurement software framework qKit. MFD, BD and AB were supported by NSERC, the Ministère de l’Économie et de l’Innovation du Québec, the Canada First Research Excellence Fund, and the U.S. Department of Energy, Office of Science, National Quantum Information Science Research Centers, Quantum Systems Accelerator.



\clearpage

\renewcommand{\theequation}{S.\arabic{equation}}
\setcounter{equation}{0}

\onecolumngrid

\section*{SUPPLEMENTAL MATERIAL}

\section{Measurement setup}
\label{A_SetUp}

\Cref{fig:setup} shows the setup used for all the qubit measurements. The samples are measured in a copper waveguide setup similar to Ref.~\cite{Maleeva_2018}, which is then encased in a shielding barrel with eccosorb glue on a copper cylinder, an Aluminium cylinder, and a Cryoperm shield. The readout line is connected to the bottom of the waveguide, which acts as a Purcell filter below the cutoff frequency of 6 GHz. There is then a pin closer to the device which is used to drive and offset the qubit. Device A exhibits a strong variation of dephasing time $T_2$ with gate charge, which is linked to the charge dispersion and noise coming from the qubit drive pin. After shortening the pin and moving the device away from the qubit drive, the measurement reported in \cref{fig_matchings} was done such that 2e is around 40 mV and the $T_2$ value vs gate charge indicates a charge noise of 0.02 in units of 2e \bd{with a maximum of $\SI{20}{\mathrm{\micro s}}$. Device B has a fixed value of $T_2$ of $\SI{14}{\mathrm{\micro s}}$.} We measure the voltage of 2e at 80 mV. We amplify the signal using a Josephson amplifier \cite{DJJA_ref} with an amplification between 15 and 20 dB for both devices. The signal is then reamplified at 4K by the HEMT amplifier and at room temperature using an LNF LNF-LNR1\textunderscore15B\textunderscore SV amplifier. For device B measurements, a homemade RF filter with a bandwidth of 30 MHz was added to the readout driving line to remove some unwanted sidebands of the drive pulses. The microwave signals are sent and recorded using a Presto machine, a frequency signal generation and analysis platform from Intermodulation Products AB.

\begin{suppfigure}[!t]
\includegraphics[width = 0.5\columnwidth]{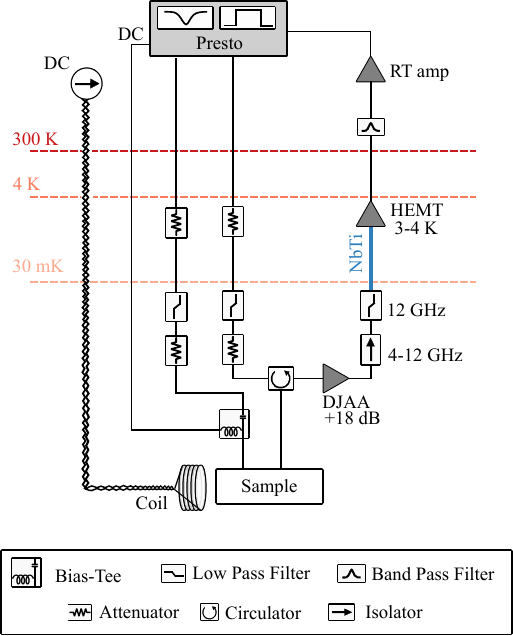}
\caption{\textbf{ Experimental setup } }
\label{fig:setup}
\end{suppfigure}

\section{Semiclassical model}
\label{A_semi_classical}

This section outlines the derivation of \cref{eq:H_driven_transmon} in the main text. The static transmon-resonator system Hamiltonian is
\begin{align}
    \label{eq:H_transmon_resonator}
    \hat{H}_{tr}=\hat H_t+ \omega_r \hat a ^\dag \hat a-i g(\hat{n}_t-n_g)(\hat{a}-\hat{a}^\dag),
\end{align}
with $\hat{H}_t$ the transmon Hamiltonian given by \cref{eq:H_transmon_main} of the main text, $\hat{n}_t$ the transmon charge operator, $\hat a$ the annihilation operator of the resonator, $\omega_r$ the bare resonator frequency, and $g$ the coupling strength. The system parameters are obtained from spectroscopy as detailed in \cref{A_higher_harmonics_fit}. In the presence of a drive on the resonator, the full Hamiltonian is 
\begin{align} 
    \label{eq:H_transmon_resonator_drive}
    \hat H(t)=\hat{H}_{tr}-i\varepsilon_d\sin(\omega_d t)(\hat{a}-\hat{a}^\dag) ,
\end{align}
where $\varepsilon_d$ and $\omega_d$ are the drive amplitude and frequency. In a frame rotating at frequency $\omega_d$ and neglecting the fast-rotating terms, the Hamiltonian is 
\begin{align}
    \label{eq:H_transmon_resonator_drive_RWA}
    \hat H(t)=&\hat H_t+ (\omega_r-\omega_d) \hat a ^\dag \hat a-i g(\hat{n}_t-n_g)(\hat{a}e^{-i\omega_d t}-\hat{a}^\dag e^{i\omega_d t}) -i\varepsilon_d(\hat{a}e^{-i\omega_d t}-\hat{a}^\dag e^{i\omega_d t}).
\end{align}
We apply a displacement transformation on the resonator which results in the replacement
\begin{align}
    \hat{a} \rightarrow \hat{a} + \alpha(t),
\end{align}
where $\alpha(t)$ is the coherent state amplitude and the remaining $\hat a$ on the right-hand-side denotes quantum fluctuations of the resonator. 
The semiclassical approximation neglects the quantum fluctuations, leading to the driven transmon Hamiltonian
\begin{align}
    \label{eq:H_transmon_resonator_drive_final}
    \hat H(t)=\hat H_t + 2g\sqrt{\bar{n}_r(t)} \cos[\omega_d t+\phi(t)](\hat{n}_t-n_g).
\end{align}
Here, $\bar{n}_r(t)=|\alpha(t)|^2$ is the average photon number and $\phi(t)$ is a slowly oscillating phase of the resonator field. Since it varies slowly on the timescale of a resonance crossing, this phase can be neglected---it does not affect the Landau-Zener dynamics responsible for population transfer~\cite{Shillito2022Dynamics,dumas2024unified}. \bd{This yields \cref{eq:H_driven_transmon} of the main text, which is used for all calculations.}

In the dispersive readout, the resonator field takes a value $\alpha_i(t)$ that is conditional on the transmon state $i$.
The semiclassical dynamics of $\alpha_i(t)$ is 
determined by the associated photon-number-dependent pulled resonator frequency $\tilde{\omega}_{r,i}(|\alpha|^2)$ for the $i$th state with the equation of motion \cite{Shillito2022Dynamics}
\begin{align}
    \dot{\alpha}_{i}=-i[\tilde{\omega}_{r,i}(|\alpha_{i}|^2)-\omega_d]\alpha_{i} -\kappa\alpha_{i}/2 -i\varepsilon_d/2.
    \label{eq:driven_damped_osc}
\end{align}
We discuss how to extract the functional form of $\tilde{\omega}_{r,i}(|\alpha|^2)$ in \cref{A_Floquet}. Our dynamical simulations are performed by first solving \cref{eq:driven_damped_osc} and then using the result to solve the Schr{\"o}dinger equation under \cref{eq:H_transmon_resonator_drive}.

Importantly, when part of the qubit population transitions to another state under a multiphoton resonance, the system evolves into an entangled transmon-resonator state that involves multiple $\alpha_i(t)$. The resonator state then becomes highly nonclassical, highlighting a key limitation of the semiclassical treatment. However, \bd{\cref{eq:driven_damped_osc} can still be used to obtain the resonator evolution associated with each distinct transmon state~\cite{Shillito2022Dynamics}.} \bd{In this work, we choose the $\tilde{\omega}_{r,i}(|\alpha|^2)$} associated with the state population that remains in the initial qubit state throughout the readout process; see \cref{A_Floquet}. This allows us to investigate the dynamics of measurement-induced transitions at multiple critical photon numbers for a given initial qubit state.

\section{Parameter fit with higher-order harmonics}
\label{A_higher_harmonics_fit}

The transmon Hamiltonian for a multiharmonic Josephson potential is given by \cref{eq:H_transmon_main} in the main text. It has been shown that higher-order harmonics, $E_{Jm}$ for $m>1$, must be considered to accurately describe the energies and charge dispersion of the transmon states beyond the qubit subspace~\cite{Willsch2024higherharmonics}. 

The parameters $(E_C,E_{J1},E_{J2},E_{J3},g,\omega_r)$ of \cref{eq:H_transmon_resonator} for device B are fitted to spectroscopy data by minimizing the loss function 
\begin{align}
    \label{eq:minimized_fun}
    f=\sum_{i=1}^3 \frac{1}{i}|\omega_{0i,n_g=0}^{\mathrm{model}}-\omega_{0i,n_g=0}^{\mathrm{exp}}| + \sum_{i=2}^3 \frac{1}{i}|\omega_{0i,n_g=0.5}^{\mathrm{model}}-\omega_{0i,n_g=0.5}^{\mathrm{exp}}| + \sum_{i=0}^1|\omega_{r,i}^{\mathrm{model}}-\omega_{r,i}^{\mathrm{exp}}|
\end{align}
using the Sequential Least Squares Programming optimization algorithm. Here, the $\omega_{0i}$ are the qubit transition frequencies between states $0$ and $i$, and $\omega_{r,i}$ is the pulled resonator frequency corresponding to transmon state $i$. The model frequencies, $\omega_{0i}^{\mathrm{model}}$ and $\omega_{r,i}^{\mathrm{model}}$, are obtained from numerical diagonalization of \cref{eq:H_transmon_resonator}. The experimental qubit transition frequencies, $\omega^{\mathrm{exp}}_{0i}$, \bd{are measured in a Ramsey experiment where we set $n_g=0$. Because of rapid quasiparticle tunneling events, the Ramsey signal contains the qubit frequencies for $n_g=0$ and $n_g=0.5$, both of which are included in the loss function. However, note that we only include the transition frequency $\omega_{10}$ at $n_g=0$ because of its negligible charge dispersion.} The experimental pulled resonator frequencies, $\omega^{\mathrm{exp}}_{r,i}$, are measured from the resonator response. The factor $1/i$ ensures that a larger weight is given to transitions involving lower-energy transmon states, since the quantities $\omega_{0i}^{\mathrm{exp}}/i$ are all roughly measured to a precision of $\sim\SI{0.1}{MHz}$. To avoid overfitting \bd{the multiharmonic model}, harmonics $E_{Jm}$ of order $m \geq 4$ are set to zero, ensuring that the number of fitted parameters remains smaller than the number of measured frequencies. \bd{For a similar reason, only the first and second transmon transitions are included in the loss function when fitting the conventional single-harmonic model.}

The resulting fit parameters are shown in \Cref{tab:parameters}. The first row presents the parameters fitted to the multiharmonic model. The rapid decay of the Josephson energies with increasing index $m$
justifies the approximation of neglecting harmonics with $m \ge 4$. The second row presents the parameters fitted to the conventional single-harmonic transmon model. Significant differences in the fit parameters are observed between the two models, notably a change in $E_J/E_C$ from 40.2 to 43.5. 
\begin{supptable*}
    \setlength{\tabcolsep}{6pt}
    \begin{center}
    \begin{tabular}{ c | c c c c c c c } 
     \hline
     \hline
     Model & $E_C/2\pi$ (MHz) & $E_{J1}/2\pi$ (GHz) & $E_{J1}/E_C$ & $E_{J2}/E_{J1}$ & $E_{J3}/E_{J1}$ & $\omega_r/2\pi$ (GHz) & $g/2\pi$ (MHz)\\ 
     \hline
     Multiharmonic & 216.6 & 8.718 & 40.2 & -0.768 \% & 0.0398 \% & 7.04767 & 186.5\\ 
     Conventional & 205.6 & 8.948 & 43.5 & 0 & 0 & 7.04765 & 181.9 \\ 
     Stray inductance &217.4 & 8.694 & 40.0 & -0.765 \% & 0.0117 \% & 7.04805 & 180.7 \\ 
     \hline
    \hline
    \end{tabular}
    \end{center}
    \caption{Fitted parameters for device B.}
    \label{tab:parameters}
\end{supptable*}

To investigate the origin of the higher-order harmonics in our device, we fit the $E_{Jm}$ to a model in which they arise from a stray inductor with inductive energy $E_L$ in series with the Josephson junction; see the Supplementary Information of \bd{Refs.}~\cite{Willsch2024higherharmonics,PhysRevApplied.22.044063}.
In the limit where $E_J/E_L\ll 1$, the higher harmonics $E_{Jm}$ are related to the Josephson and inductive energies by 
\begin{align}
    \begin{aligned}
    &E_{J1}\approx E_J\Bigg[1-\frac{1}{8}\Big(\frac{E_J}{E_L}\Big)^2+\frac{1}{192}\Big(\frac{E_J}{E_L}\Big)^4\Bigg] ,\\
    &E_{J2}\approx E_J\Bigg[-\frac{1}{4}\Big(\frac{E_J}{E_L}\Big)+\frac{1}{12}\Big(\frac{E_J}{E_L}\Big)^3-\frac{1}{96}\Big(\frac{E_J}{E_L}\Big)^5\Bigg] ,\\
    &E_{J3}\approx E_J\Bigg[\frac{1}{8}\Big(\frac{E_J}{E_L}\Big)^2-\frac{9}{128}\Big(\frac{E_J}{E_L}\Big)^4\Bigg].
    \end{aligned}
\end{align}
Fitting our data to this model yields $E_J/2\pi=\SI{8.693}{GHz}$ and $E_L/2\pi=\SI{284.2}{GHz}$. \bd{This corresponds to a linear inductance of $L=\SI{0.575}{nH}$ which is consistent with that device's geometry (see supplementary material of Ref.~\cite{Willsch2024higherharmonics}).} The third row of~\cref{tab:parameters} shows the values of the higher-order harmonics and of the other parameters resulting from this fit. We note that the obtained parameters are in close range to the multiharmonic model ones, as is further discussed in \cref{A_higher_harmonics_impact}.

\section{Floquet branch analysis}
\label{A_Floquet}

This section describes the Floquet branch analysis, a procedure for the identification of the critical photon numbers at which population is expected to leak outside the qubit subspace during readout~\cite{dumas2024unified}. 

When $\kappa\ll\omega_d$, the effective drive amplitude $\varepsilon_t(t)=2g\sqrt{\bar n_r(t)}$ in \cref{eq:H_driven_transmon} varies on a timescale much longer than the period of the drive, $T=2\pi/\omega_d$. Thus, the Hamiltonian is approximately periodic at any given time, yielding an instantaneous Floquet spectrum at that time~\cite{Breuer1989QuantumPhases}. \bd{For a given $\varepsilon_t$,} the Floquet modes $\ket{\phi[\varepsilon_t]}$ and quasienergies $\epsilon[\varepsilon_t]$ are obtained by diagonalizing the one-period propagator
\begin{align} 
    \hat{U}(t+T,t)\ket{\phi(t)}=e^{-i\epsilon T}\ket{\phi(t)}.
    \label{eq:propagator}
\end{align}
The Floquet branch analysis classifies the modes and quasienergies into Floquet branches. At zero drive amplitude, the Floquet modes are just the bare transmon states, $\ket{\phi_{i}[0]}=\ket{i}$.
The Floquet branch $B_{i}$ associated with state $\ket{i}$ is constructed from the bare states as follows. The drive amplitude is progressively increased in increments of $\delta\varepsilon_t=\SI{5}{MHz}$. For each amplitude $\varepsilon_t$, the next Floquet mode $\ket{\phi [\varepsilon_t]}$ and associated quasienergy $\epsilon [\varepsilon_t]$ of the branch are chosen by maximizing the overlap 
\begin{align}
    \big|\big\langle\phi[\varepsilon_t]\big|\phi_{i}[\varepsilon_t-\delta\varepsilon_t]\big\rangle\big|^2.
\end{align}
The resulting Floquet branches for the multiharmonic model parameters of device B \bd{(see \cref{A_higher_harmonics_fit})} are shown in \cref{fig:floquet_BA}. Panel (a) shows the quasienergies of the Floquet branches, and panel (b) shows the average transmon population of the Floquet modes, $N_{t,i}=\sum_{j}j|\langle{j}|\phi_{i}(0)\rangle|^2$. Both quantities are plotted as a function of the average photon number $\bar{n}_r = (\varepsilon_t/2g)^2$. At specific photon numbers, resonances between the quasienergies of different branches lead to significant hybridization of the Floquet modes. This results in avoided crossings of the quasienergies in panel (a) and in a swapping of the transmon populations of the Floquet branches in panel (b). The avoided crossings of the ground and excited state branches allow us to identify the Floquet critical photon numbers for readout, which are indicated by the dotted vertical lines.
\begin{suppfigure}[!t]
\includegraphics[width = 0.7\columnwidth]{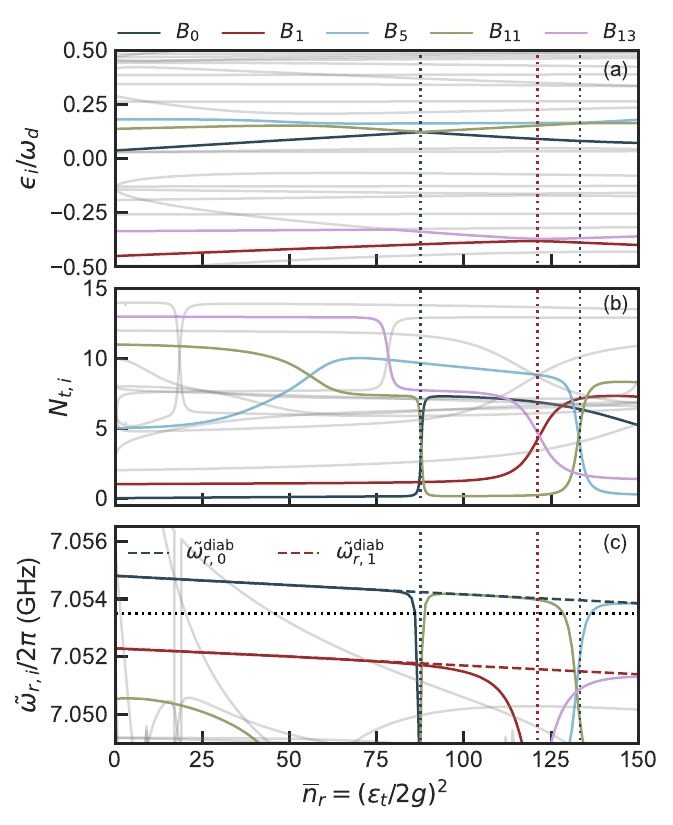}
\caption{\textbf{Floquet branch analysis of device B.} (a) Floquet quasienergies, (b) average population of the Floquet modes, and (c) photon-number-dependent resonator frequency $\tilde{\omega}_{r,i}(\bar{n}_r)$ as a function of the average photon number $\bar{n}_r=(\varepsilon_t/2g)^2$ for $n_g=0.23$. The dotted vertical lines indicate multiphoton resonances for the ground (blue) and excited (red) qubit states. In (c), $\tilde\omega_{r,0}^{\mathrm{diab}}$ (blue dashed line) and $\tilde\omega_{r,1}^{\mathrm{diab}}$ (red dashed line) are the pulled resonator frequencies obtained by diabatically tracking the qubit branches. The dotted black line indicates the value of $\omega_d$. Branches $B_{0}$, $B_{1}$, $B_{5}$, $B_{11}$, and $B_{13}$ are highlighted in color.}
\label{fig:floquet_BA}
\end{suppfigure}

Because the increment $\delta \varepsilon_t$ is chosen to be relatively small, the tracking emulates an adiabatic increase in the photon number $\bar{n}_r$ as a function of time. Such an adiabatic evolution through an avoided crossing results in full population transfer at that avoided crossing. Here, we aim to track the qubit population \bd{\emph{diabatically}} beyond the first crossing to capture multiple critical photon numbers. In \cref{fig:floquet_BA}(b), for example, the population initialized in $B_{0}$ can transition to higher-energy levels at $\bar{n}_r\approx 88$. However, if the population diabatically transfers to $B_{11}$, which is $0$-like after the crossing, the next critical photon number is found at $\bar{n}_r \approx 133$ due to an avoided crossing with $B_{5}$. \bd{The avoided crossings obtained from the diabatic tracking are the ones shown in \cref{fig:readout_demolition_comparison}(b).}

Even though the semiclassical model does not explicitly include the resonator, the Floquet analysis nevertheless enables the computation of the pulled resonator frequencies for each qubit state and for each resonator photon number. \bd{Indeed, we can} assign to each Floquet branch $B_{i}$ an effective nonlinear oscillator with photon-number dependent frequency $\tilde{\omega}_{r,i}(\bar{n}_r)$ given by
\begin{align}
    \tilde{\omega}_{r,i}(\bar{n}_r)=\omega_r+\epsilon_{i}(\bar{n}_r+1)-\epsilon_{i}(\bar{n}_r),
\end{align}
where $\epsilon_{i}(\bar{n}_r)$ is the Floquet quasienergy of branch $B_{i}$ computed at the effective drive amplitude $\varepsilon_t=2g\sqrt{\bar{n}_r}$. The frequencies obtained in this way are very close to those obtained from the diagonalization of the full transmon-resonator Hamiltonian~\cite{dumas2024unified}. This method thus accounts for nonlinear effects at high photon numbers while remaining computationally more efficient than full diagonalization. The effective oscillator frequencies of the Floquet branches are shown in \cref{fig:floquet_BA}(c). They display significant nonlinear photon number dependence around the critical photon numbers. To dynamically follow the population in a qubit-like branch, we define $\tilde{\omega}_{r,i}^{\mathrm{diab}}$, the effective oscillator frequency obtained by diabatically tracking branch $B_{i}$ \bd{through the avoided crossings} as described above.
\bd{Away from the avoided crossings,} the pulled resonator frequencies $\tilde{\omega}^\mathrm{diab}_{r,0}$ and $\tilde{\omega}^\mathrm{diab}_{r,1}$ vary approximately linearly with $\bar{n}_r$, with slopes $K_{0}/2\pi\approx-\SI{6.6}{kHz}$ and $K_{1}/2\pi\approx-\SI{5.5}{kHz}$ corresponding to the resonator self-Kerr nonlinearities. \bd{At the avoided crossings, we interpolate between these nearly linear regions for simplicity.} The diabatic oscillator frequencies \bd{obtained in this manner are shown as dashed colored lines in panel (c)} for branches $B_{0}$ and $B_{1}$.

The choice to follow the resonator dynamics associated with a qubit-like branch stands on the assumption that most of the state population \bd{mostly} remains in the initial branch for the whole readout duration. However, this approximation does not always hold. For example, at large photon numbers, measurement-induced transitions can sometimes lead to population transfers exceeding 50\% due to the presence of \bd{wide avoided crossings} in the quasienergy spectrum. Despite this limitation, this choice allows us to model the resonator dynamics for the remaining qubit population. \bd{Thus,} the time dynamics simulations correctly identify many photon numbers at which we observe drops in the experimental qubit state survival probability \bd{beyond the first threshold}. However, with this approach, we expect the modeling of the resonator dynamics to be inaccurate for the state population transferred to high-energy states, for which the photon-number dependent frequency $\tilde{\omega}_{r,i}$ can be significantly off-resonant from that of the qubit-like branch, $\tilde{\omega}_{r,i}^{\mathrm{diab}}$. Therefore, we do not expect the simulations to quantitatively reproduce the population transfer from higher-energy states to the qubit-like state during the resonator ramp-down. In \cref{fig:readout_demolition_comparison}, \bd{we believe that part of the discrepancy} between the qubit state survival probability for the experimental data in panel (b) and the time dynamics simulations in panel (c) \bd{can be} attributed to this source of error.

\section{Nonlinear photon number calibration}
\label{A_Photon_calib}

Here, we describe the procedure used to calibrate the photon number as a function of the \bd{externally applied MW drive} voltage \bd{for device B}. At low photon numbers, the qubit frequency is ac-Stark shifted proportionally to the average resonator photon number $\bar n_r$,
\begin{align}
    \label{eq:qubit_shift}
    \tilde{\omega}_q(\bar n_r) = \omega_q + \chi \bar n_r.
\end{align}
Here, $\chi= \chi_1 - \chi_0$ is the full dispersive shift, with $\chi_0$ and $\chi_1$ the dispersive shifts for the ground and excited states, respectively. \bd{We first measure $\tilde{\omega}_q$ as a function of the MW drive voltage applied to the resonator in the linear regime. We then convert $\tilde{\omega}_q$ into a photon number $\bar{n}_r$ using the measured $\chi$ shift and \cref{eq:qubit_shift}.} \Cref{fig:photon_calibration}(a) shows the experimentally measured qubit frequency as a function of the applied \bd{MW drive} voltage (purple dots). The values of $\chi_0$ and $\chi_1$ are measured separately and subtracted to yield the total dispersive shift $\chi/2\pi =\SI{-2.5}{MHz}$. This value yields the relationship between photon number and applied \bd{MW drive} voltage shown as the purple dots in \cref{fig:photon_calibration}(b).

\begin{suppfigure}[!t]
\includegraphics[width = 0.8\columnwidth]{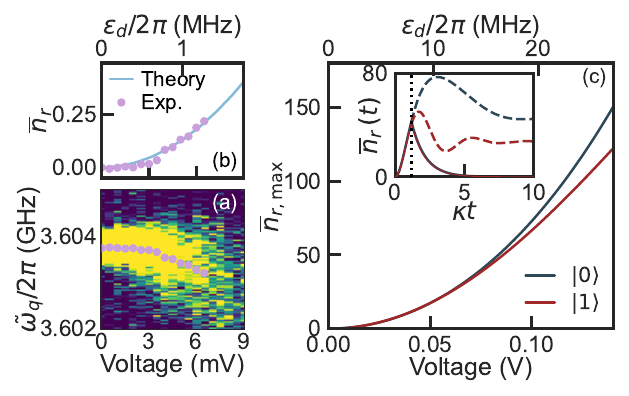}
\caption{\textbf{Photon number calibration for device B. }{(a) Experimentally measured qubit frequencies as a function of applied \bd{MW drive} voltage (purple dots). The background shows the measured spectroscopic response for each \bd{MW drive} voltage. (b) Steady-state photon number extracted from \cref{eq:qubit_shift} as a function of \bd{MW drive} voltage (purple dots), and computed by solving \cref{eq:driven_damped_osc} as a function of the resonator drive amplitude $\varepsilon_d$ (light blue line). The curves \bd{are matched} by rescaling the applied \bd{MW drive} voltage, \bd{showing that $\varepsilon_d$ is directly proportional to the MW drive voltage.} (c) Maximum average photon number \bd{$\bar{n}_{r,{\rm max}}$} obtained by solving \cref{eq:driven_damped_osc} as a function of applied \bd{MW drive} voltage for a pulse duration of 200 ns when the qubit is initialized in the ground state (blue) or excited state (red). The inset shows \bd{time-dependent photon number $\bar n_r(t)$} for $\varepsilon_d/2\pi=\SI{15}{MHz}$. The solid lines show \bd{$\bar{n}_r(t)$} when the drive is turned off after 200 ns (dotted vertical line) for the qubit initialized in the ground state (blue) or excited state (red). The dashed lines show \bd{$\bar{n}_r(t)$} if the drive is not turned off after 200 ns.}}
\label{fig:photon_calibration}
\end{suppfigure}

\bd{At larger photon numbers, the above linear calibration is not expected to be accurate \cite{Gambetta2006}. Nevertheless, we now show that it can be leveraged to calibrate the resonator drive $\varepsilon_d$. Once $\varepsilon_d$ is known, solving \cref{eq:driven_damped_osc} yields the average photon number $\bar{n}_r(t)$ for any MW drive voltage. To calibrate $\varepsilon_d$, we compute the steady-state photon number when the qubit is in the ground state, $\bar{n}_{r}\equiv|\alpha_{0}|^2$, as a function of $\varepsilon_d$. This is done by solving \cref{eq:driven_damped_osc} using the experimental drive frequency $\omega_d$ and the (diabatic) nonlinear dispersion $\tilde{\omega}_{r,0}^{\mathrm{diab}}(\bar{n}_r)$ for $i=0$ (see \cref{A_Floquet}). The result is shown in \cref{fig:photon_calibration}(b). Rescaling the externally applied MW drive voltage allows us to match photon numbers obtained from linear calibration (purple dots) to the calculated steady-state values (light blue line). This establishes a direct proportionality relation between the MW drive voltage and $\varepsilon_d$. We assume that this relation holds for the larger range of MW drive voltages used in this experiment because the measurement chain is expected to have a linear response in that range.}

\Cref{fig:photon_calibration}(c) shows the maximum photon number \bd{$\bar{n}_{r,{\rm max}}$ reached} for a pulse duration of 200~ns as a function of the applied \bd{MW drive} voltage. The blue line shows the result when the qubit is initialized in the ground state with dispersion $\tilde{\omega}_{r,0}^{\mathrm{diab}}(\bar{n}_r)$, and the red line shows the result when the qubit is initialized in the excited state with dispersion $\tilde{\omega}_{r,1}^{\mathrm{diab}}(\bar{n}_r)$. The upper axis shows the corresponding drive amplitude $\varepsilon_d$. At large \bd{MW drive} voltages, the resonator photon number reaches considerably larger values when the qubit is in the ground state. This occurs because the drive frequency $\omega_d$ is set between the pulled resonator frequencies for the two qubit states and because the self-Kerr nonlinearity is negative. For that reason, the frequency $\tilde{\omega}_{r,0}^{\mathrm{diab}}(\bar{n}_r)$ becomes more resonant with $\omega_d$ at large photon numbers; see \cref{fig:floquet_BA}(c). By contrast, $\tilde{\omega}_{r,1}^{\mathrm{diab}}(\bar{n}_r)$ becomes more off-resonant with $\omega_d$ at large photon numbers. The inset of \cref{fig:photon_calibration}(c) shows the time-dependent photon number for $\varepsilon_d/2\pi=\SI{15}{MHz}$ when the qubit is initialized in the ground state (blue line) or in the excited state (red line). The vertical dotted line indicates the experimental pulse duration $t=\SI{200}{ns}$. We also show what the evolution of the photon number would be for the ground state (dashed blue line) and for the excited state (dashed red line) if the drive were not turned off beyond $200$ ns. Since $\kappa\sim|\omega^\mathrm{diab}_{r,i}-\omega_d|$ for both initial qubit states, the oscillator is in the underdamped regime and the photon number oscillates in time. Nevertheless, there are no oscillations in the photon number within the experimental readout pulse duration of 200 ns. We note that this regime is different from the one investigated in Ref.~\cite{dumas2024unified}, where it was assumed that $\kappa\gg|\omega^\mathrm{diab}_{r,i}-\omega_d|$. Calibrating the photon number in the appropriate regime is crucial to accurately model the resonator dynamics and identify the critical photon number.

Finally, we note that we only calibrate the photon number for one value of $n_g$ since the diabatic dispersions $\tilde{\omega}_{r,i}^{\mathrm{diab}}(\bar{n}_r)$ is largely insensitive to $n_g$ for the computational states. These dispersions also ignore the effect of the $n_g$-sensitive multiphoton resonances (see \cref{A_Floquet}). Therefore, we expect the calibration to be approximately the same for all $n_g$.

\section{Dynamics of the driven transmon}
\label{A_dynamics}

In this section, we describe the time dynamics simulations of the driven transmon used to study the dynamics of qubit ionization in device B. The transmon state is first initialized in the ground or excited state, $\ket{\psi(t_0)}=\ket{i}$ with $\ket{i}=\ket{0}$ or $\ket{1}$, respectively. It is then evolved by solving the Schrödinger equation for the effective time-dependent Hamiltonian in \cref{eq:H_driven_transmon}, with solution at time $t$ given by $\ket{\psi_i(t)}=\hat U(t,t_0)\ket{i}$. The time-dependent average photon number $\bar{n}_r(t)=|\alpha_{i}(t)|^2$, which determines the effective drive amplitude $\varepsilon_t(t)=2g\sqrt{\bar{n}_r(t)}$ in \cref{eq:H_driven_transmon}, is computed from \cref{eq:driven_damped_osc} with initial qubit state $i$. The resonator drive $\varepsilon_d$ is switched on at the initial time, leading to a gradual increase of $\bar{n}_r(t)$. The drive is then turned off after $\SI{200}{ns}$, after which the resonator gradually empties on a timescale of $\SI{1}{\mu s} \approx 6/\kappa$. We compute the Floquet basis $\ket{\phi_{i}[\varepsilon_t(t)]}$ associated with the instantaneous value of $\varepsilon_t(t)$. 
The projection of the time-evolved state on that instantaneous Floquet basis then gives the time-dependent probability $P(j|i)$ of transitioning from the initial $i$th branch to the final $j$th branch:
\begin{align}
    P(j|i)= |\langle \phi _{j}[\varepsilon_t(t)]|\psi_i(t)\rangle|^2.
\label{eq:overlap_evolution}
\end{align}
In particular, the quantity $1-P(i|i)$ describes the final population transfer out of the initial Floquet branch at the final time $t=\SI{1.2}{\mu s}$ and is thus a proxy for the experimentally observed probability of measurement-induced transitions (assuming that the low-power readout pulses \bd{used for preparation and readout} cause negligible state transitions).

Whenever the photon number $\bar{n}_r$ approaches an avoided crossing in the Floquet quasienergies, the population generically splits between the two associated Floquet branches. These populations subsequently interfere in a way that affects the population transfer at the final time. Experimentally, however, rapid decoherence of such superpositions is expected since the resonator pulls $\tilde\omega_{r}$ of the Floquet branches are substantially different; see \cref{fig:floquet_BA}(c). Any fluctuation in the photon number due to, e.g., quantum fluctuations, leads to fluctuations in the quasienergy difference between the branches and thus to dephasing. The aforementioned interference effects, which are a feature of our fully coherent semiclassical model, are thus not expected to be observed in practice. To roughly mimic this unavoidable dephasing between Floquet branches, the coherences in the Floquet basis are manually set to zero at the end of the ramp up at $t=\SI{200}{ns}$. The diagonal elements of this decohered state are then evolved separately and coherently during the ramp down. The final transition probability is reconstructed by summing the transition probabilities for all possibilities:
\begin{align}
    P(i|i)= \sum_{j} P_\mathrm{down}(i|j)P_\mathrm{up}(j|i).
\label{eq:overlap_evolution_2}
\end{align}
Here, $P_\mathrm{up}(j|i)=|\langle \phi_{j}[\varepsilon_t(t_\textrm{up})]|\psi_i(t_\textrm{up})\rangle|^2$ is the population transferred from the prepared $i$th transmon state to the $j$th Floquet mode after a ramp-up of $t_{\textrm{up}}=\SI{200}{ns}$. The quantity $P_\mathrm{down}(i|j)=|\langle \phi_{i}[\varepsilon_t(t_f)]|\hat U(t_f,t_\mathrm{up})|\phi_{j}[\varepsilon_t(t_\mathrm{up})]\rangle|^2$ is the transition probability from the state initialized in the $j$th Floquet mode at the time $t_\textrm{up}$ back to the $i$th Floquet mode at the end of the ramp-down at the final time $t_{f}=1.2~\mu$s. In the bottom row of \cref{fig:readout_demolition_comparison}(b) of the main text, we plot the quantity $1 - P(i|i)$ computed from \cref{eq:overlap_evolution_2} for various values of the drive amplitude $\varepsilon_d$ and of the offset charge $n_g$.

To highlight the dynamical nature of the transition process, we plot the transition probability as a function of time in \cref{fig:image_sc_vs_quantum}(a) for \bd{$n_g=\{0.42,0.45,0.48\}$;} see the solid lines. \Cref{fig:image_sc_vs_quantum}(b) shows the average photon number as a function of time, with the horizontal dashed lines showing the critical photon numbers obtained from the Floquet analysis for the same values of $n_g$ as in (a). The results highlight the importance of considering the effect of both the photon number ramp-up and ramp-down on the final transition probability, with sharp variations of the probability observed when the critical photon number is reached during both phases. The qubit population transitions at later or earlier times depending on the charge value, which is due to resonances occurring at critical photon numbers that depend on $n_g$. Note that the probabilities always peak near 50\% at the critical photon numbers, as the Floquet modes are in an equal superposition of the qubit states at that point.
\begin{suppfigure}[!t]
\includegraphics[width = 0.7\columnwidth]{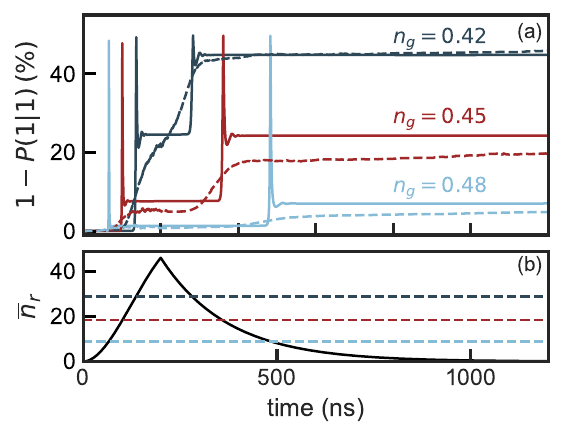}
\caption{\textbf{Temporal profile of measurement-induced transitions in simulations of device B. }{(a) Probability to transition out of the qubit excited state branch as a function of time when solving the dynamics of the semiclassical model (solid lines) and of the quantum model (dashed lines) for three different values of the gate charge. (b) Average photon number calculated from the semiclassical model for a drive amplitude $\varepsilon_d/2\pi=\SI{16}{MHz}$ turned on for $\SI{200}{ns}$ when initializing the qubit in the excited state. The dashed lines correspond to the Floquet critical photon numbers for the same values of the gate charge as in (a).}}
\label{fig:image_sc_vs_quantum}
\end{suppfigure}

\bd{We further validate} the semiclassical model by computing the time dynamics of the full quantum model described by \cref{eq:H_transmon_resonator}, which evolves under the following Lindblad master equation,
\begin{align}
    \label{eq:lindblad_eq}
    \partial_t \hat{\rho} =
    -i[\hat{H}_{tr} -i\varepsilon_d\sin(\omega_d t)(\hat{a}-\hat{a}^\dag) , \hat{\rho}] + \kappa \mathcal{D} [\hat a] \hat{\rho}.
\end{align}
Here, $\varepsilon_d$ and $\omega_d$ are the drive amplitude and frequency, $\kappa$ is the single-photon loss rate of the resonator, and $\mathcal{D} [\hat a]\hat{\rho}$ is the usual Lindblad dissipator. \bd{We solve the Lindblad eqation using a Monte-Carlo solver with} 500 quantum trajectories for $\varepsilon_d/2\pi=\SI{16}{MHz}$ and the same three values of \bd{$n_g$} as in \cref{fig:image_sc_vs_quantum}(a). The state is initialized in the dressed excited qubit state at zero photon number.  \bd{At each time, we calculate the population of all states obtained by diabatic tracking of the excited state. These states are identified using a diabatic version of the branch analysis of Ref.~\cite{Shillito2022Dynamics}.} \bd{This probability is directly analogous to the one computed using \cref{eq:overlap_evolution} and \cref{eq:overlap_evolution_2}} for the semiclassical dynamics. The results are shown as a function of time in \cref{fig:image_sc_vs_quantum}(a). \bd{For all three values of $n_g$,} the quantum time dynamics show that a sharp increase in state transition probability occurs at the same photon numbers and times as in the semiclassical model. Furthermore, both models \bd{show a similar} increase in population transfer around the critical photon number during both \bd{ramp-up and ramp-down.} \bd{This increase} is smoother for the quantum model than for the semiclassical one due to the quantum photon number fluctuations in the resonator. Furthermore, relatively good agreement in the final transition probability is observed between \bd{the two models.} The agreement between the results of both models validates the use of the semiclassical model to predict both the onset and the probability of measurement-induced transitions, with the added benefit of being much more computationally efficient.

\section{Impacts of charge-sensitive states on measurement-induced transitions}
\label{A_deep_transmon}

\bd{In this section, we show that even deeper} in the transmon regime, the onset of measurement-induced transitions is expected to remain charge-dependent. Indeed, while the charge sensitivity of the qubit states decreases exponentially with $E_J/E_C$, the higher-energy levels involved in the resonances do not benefit from this protection. \Cref{fig:deep_transmon}(a) shows the charge-dependent Floquet critical photon numbers as a function of readout frequency for a deep transmon with $E_J/E_C=100$. The critical photon numbers are shown as colored dots for both the ground and excited states, with the colors \bd{indicating} six different values of the gate charge. The area of the dots is proportional to the size of the quasienergy gap. Strong dependence of the position of the resonances on the gate charge is observed. In some range of $\omega_d$, the critical photon numbers are on average larger. However, there are still resonances occurring at lower photon numbers for specific values of $n_g$, even though we consider only six values of $n_g$. Therefore, selecting a readout frequency for which variations of the charge do not limit the average onset of measurement-induced transitions is very difficult, if not impossible. In contrast, \cref{fig:deep_transmon}(b) shows the same critical photon numbers as in \cref{fig:deep_transmon}(a), but for a single value of the charge, $n_g=0$. In this case, the distribution of the resonances is much less dense, and one can easily identify values of $\omega_d$ for which the onset of transitions is pushed to large photon numbers. This suggests that the critical photon numbers can on average be increased with \bd{active charge calibration} even deep in the transmon regime, which could prove beneficial to dispersive readout performance.

\begin{suppfigure}[!t]
\includegraphics[width = 0.7\columnwidth]{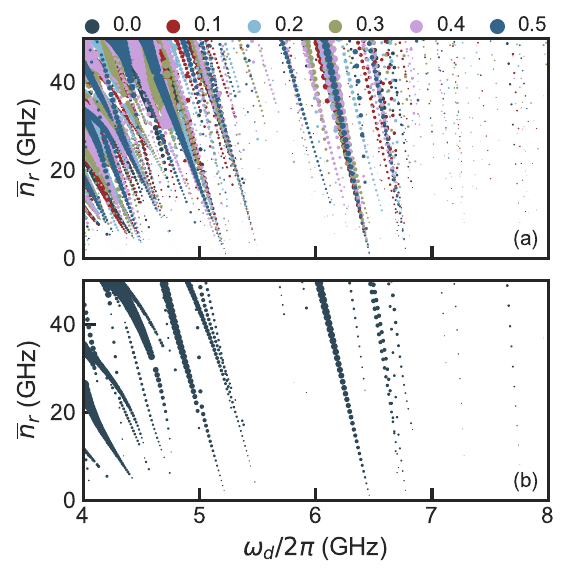}
\caption{\textbf{Measurement-induced transitions in the deep transmon regime with ${\boldsymbol{E_J/E_C=100}}$ and ${\boldsymbol{\omega_q/2\pi=}\SI{3.61}{GHz}}$.} Floquet critical photon numbers involving the qubit computational states as a function of $\omega_d$ for (a) 6 different values of offset charge linearly spaced between $n_g=0$ and $n_g=0.5$ (see legend) and (b) for $n_g=0$ only.
}
\label{fig:deep_transmon}
\end{suppfigure}

\section{Comparison of the results with and without inclusion of higher-order harmonics}
\label{A_higher_harmonics_impact}

As discussed in \cref{A_higher_harmonics_fit}, we fit both the \bd{multiharmonic model and the conventional single-harmonic model} to the measured spectroscopy data of device B \bd{(\cref{tab:parameters})}. \bd{In \cref{fig:no_harmonics}, we compare the semiclassical dynamics} \bd{of these two models for the ground state (left column) and for the excited state (right column).} Panels (a-b) reproduce the results of \cref{fig:readout_demolition_comparison}(b) in the main text; \bd{panel (a) shows} the experimental data and \bd{panel (b) shows} the semiclassical time dynamics for the parameters of the multiharmonic model. Panel (c) shows the semiclassical time dynamics \bd{for the parameters of the conventional model.} The latter do not align with the experimental data shown in (a). For example, significant transitions out of the qubit ground state is expected around $n_g\approx0$ at low photon numbers, which is not observed in the experimental data. 

To make this comparison clearer, the Floquet critical photon numbers computed from the two sets of parameters are displayed above the experimental data in panel (a). Significant mismatch is observed between the offset-charge dependent features in the experimental data and the critical photon numbers for the conventional transmon model (black circles). This is in sharp contrast with the excellent agreement between the experimental features and the critical photon numbers for the multiharmonic model (red circles). The disparity between the theoretical predictions of the two models can be explained by the significant variation in the ratio $E_{J1}/E_C$, as was discussed in Ref.~\cite{Willsch2024higherharmonics}. This ratio has a considerable impact on the energy and dispersion of the high-energy transmon states involved in measurement-induced transitions. These results support the presence of higher-order harmonics in the transmon Hamiltonian and highlight the need to carefully model the high-energy sector of the transmon, and potentially of other Josephson-junction based qubits, in order to accurately reproduce the onset of measurement-induced state transitions. 

\begin{suppfigure*}[!t]
\hspace*{0cm}
\includegraphics[width = \columnwidth]{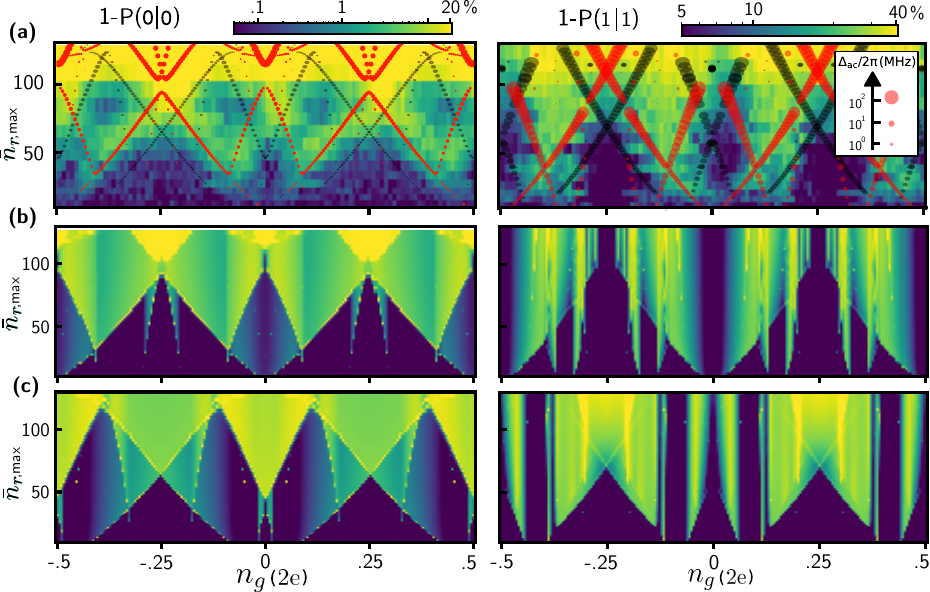}
\caption{\textbf{Impact of higher-order harmonics in the modeling of device B on measurement-induced transitions}.
Measurement-induced transitions for the ground state (left column) and the excited state (right column) as a function of offset charge and maximal readout photon number. (a) Experimental data is shown in the background, with circles on top indicating the positions of avoided crossings in the Floquet quasienergy spectrum for the multiharmonic model (red) and the conventional transmon model (black). The dot area is proportional to the gap size $\Delta_{ac}$. (b-c) Semiclassical time dynamics results using the Hamiltonian parameters fitted to (b) the multiharmonic model and (c) the conventional transmon model.}
\label{fig:no_harmonics}
\end{suppfigure*}

\bd{Note that we have verified (not shown) that} the \bd{semiclassical time dynamics for the parameters of} the series inductance fit \bd{(third row of \cref{tab:parameters})} are very similar to those shown in \cref{fig:no_harmonics}(b). This is \bd{because the parameters obtained from the series inductance fit are very similar to those obtained from the multiharmonic fit; compare the first and third rows of \cref{tab:parameters}.} The largest relative variation of the parameters resides in the third-order harmonics, $E_{J3}$. However, since $E_{J3}/E_J<0.1\%$, even relatively large variations of this parameter do not significantly impact either the onset or the probability of measurement-induced transitions. Since the experimental results can be accurately reproduced with two sets of parameters that \bd{are consistent with either additional high-transparency channels in the junction~\cite{Willsch2024higherharmonics} or a series inductance model with a realistic value of $L=\SI{0.575}{nH}$~\cite{Willsch2024higherharmonics},} our analysis does not provide further insight into the physical origin of the higher-order harmonics in the Hamiltonian. However, in the case of transmons with larger \bd{higher-order} harmonics~\cite{Willsch2024higherharmonics}, investigating the charge-dependent onset of measurement-induced transitions \bd{could provide enough experimental constraints to distinguish between various models for the origin of these higher harmonics.}

\clearpage

\bibliography{OCS_references}
\end{document}